\documentclass{aastex631}

\let\AASsplitbox\splitbox
\let\splitbox\undefined
\usepackage{adjustbox}
\let\splitbox\AASsplitbox
\usepackage{xparse}
\usepackage{xcolor}
\usepackage{amsmath} 
\usepackage{subfigure}
\usepackage{graphicx}
\usepackage{float}
\usepackage{longtable}
\usepackage{array}

\begin{document}

\title{A Massive and Eccentric Binary Millisecond Pulsar in the Globular Cluster M10}

\author[0000-0002-5344-9552]{Yujie Wang}
\affiliation{State Key Laboratory of Radio Astronomy and Technology, Shanghai Astronomical Observatory, CAS, 80 Nandan Road, Shanghai 200030, People’s Republic of China}
\affiliation{School of Astronomy and Space Sciences, University of Chinese Academy of Sciences, Beijing 100049, People’s Republic of China}

\author[0000-0002-8366-3373]{Zhongli Zhang}
\affiliation{State Key Laboratory of Radio Astronomy and Technology, Shanghai Astronomical Observatory, CAS, 80 Nandan Road, Shanghai 200030, People’s Republic of China}
\affiliation{School of Astronomy and Space Sciences, University of Chinese Academy of Sciences, Beijing 100049, People’s Republic of China}

\author[0000-0001-7771-2864]{Zhichen Pan}
\affiliation{National Astronomical Observatories, Chinese Academy of Sciences, Beijing 100010, People’s Republic of China}
\affiliation{Guizhou Radio Astronomical Observatory, Guizhou University Guiyang 550025, People’s Republic of China}
\affiliation{CAS Key Laboratory of FAST, National Astronomical Observatories, Chinese Academy of Sciences, Beijing 100101, People’s Republic of China}
\affiliation{School of Astronomy and Space Sciences, University of Chinese Academy of Sciences, Beijing 100049, People’s Republic of China}

\author[0000-0003-0597-0957]{Lei Qian}
\affiliation{National Astronomical Observatories, Chinese Academy of Sciences, Beijing 100010, People’s Republic of China}
\affiliation{Guizhou Radio Astronomical Observatory, Guizhou University Guiyang 550025, People’s Republic of China}
\affiliation{CAS Key Laboratory of FAST, National Astronomical Observatories, Chinese Academy of Sciences, Beijing 100101, People’s Republic of China}
\affiliation{School of Astronomy and Space Sciences, University of Chinese Academy of Sciences, Beijing 100049, People’s Republic of China}

\author[0000-0002-1334-8853]{Lijing Shao}
\affiliation{Kavli Institute for Astronomy and Astrophysics, Peking University, Beijing 100871, People’s Republic of China}
\affiliation{National Astronomical Observatories, Chinese Academy of Sciences, Beijing 100010, People’s Republic of China}

\author[0000-0001-6051-3420]{Dejiang Yin}
\affiliation{College of Physics, Guizhou University, Guiyang 550025, People’s Republic of China}

\author[0009-0001-6693-7555]{Yujie Lian}
\affiliation{Institute for Frontiers in Astronomy and Astrophysics, Beijing Normal University, Beijing 102206, People’s Republic of China}
\affiliation{School of Physics and Astronomy, Beijing Normal University, Beijing 100875, People’s Republic of China}

\correspondingauthor{Zhongli Zhang}
\email{zzl@shao.ac.cn}

\begin{abstract}

We present the timing solution of the millisecond pulsar PSR~J1657$-$0406A in the globular cluster M10, based on observations performed with the Five-hundred-meter Aperture Spherical radio Telescope (FAST). The timing results show that PSR J1657$-$0406A is in a long-period ($P_{\rm b} \sim 37$ days) and eccentric ($e \sim 0.12$) orbit. We also measured the periastron advance $\dot{\omega}=0.00090(13)$ deg yr$^{-1}$, which implies a total system mass of $2.5(6)\ {\rm M}_\odot$ by assuming that general relativity accurately models the orbital motion. No optical counterpart was detected within the positional uncertainty, indicating that the companion is unlikely to be a main-sequence star. Considering possible formation channels for the orbital eccentricity, we favor a helium white dwarf companion, although a massive white dwarf or neutron star cannot be completely excluded.  

\end{abstract}


\section{Introduction} \label{sec:intro}

In the studies of binary millisecond pulsars (MSPs), the orbital eccentricity plays a key role in probing their formation and evolutionary history. Typically, these systems form when a low-mass donor transfers material onto an old neutron star (NS), during which tidal forces efficiently damp the eccentricity imparted by the supernova explosion (e.g., \citealt{phinney1992pulsars}). However, unlike the commonly observed binary MSPs with nearly circular orbits, many systems exhibit significant eccentricities, thus suggesting that these eccentric systems may form through a different evolutionary path.

Due to their extremely dense stellar environments, globular clusters (GCs) are prime sites for the formation of eccentric binary MSPs. Both theoretical models and observational studies have shown that the formation of binaries and MSPs in GCs is strongly influenced by dynamical processes, such as tidal captures and exchange encounters (\citealt{hills1976formation}, \citealt{verbunt1987globular}, \citealt{verbunt2014disruption}), which can produce high-eccentricity systems on short timescales. For wide-orbit pulsar binaries with low-mass companions, their eccentricities can also gradually increase due to long-term perturbations from nearby stars \citep{rasio1995orbital}. Among the 369 pulsars discovered in 46 GCs to date, about 60\% reside in binary systems, with more than 30\% exhibiting eccentricities $>$ 0.1 \footnote{\url{https://www3.mpifr-bonn.mpg.de/staff/pfreire/GCpsr.html}}. Examples include the most eccentric binary pulsar known to date, PSR~J1835$-$3259A in NGC 6652 ($e \sim 0.97$; \citealt{decesar2015highly}); the possible MSP-NS system PSR~J1807$-$2500B in NGC 6544 ($e \sim 0.75$; \citealt{lynch2012timing}); and PSR~J0514$-$4002A in NGC 1851 ($e \sim 0.89$), which may be undergoing continuous interaction with a third body \citep{dutta2025ngc}. These large eccentricities are thought to be due to the GC environment and its associated dynamical processes.

In addition, the eccentricity of binary MSPs' orbits can also arise from intrinsic dynamical processes within the primordially formed systems themselves. In fact, a peculiar population of MSP-He white dwarf (HeWD) binaries is observed in the Galactic field, with orbital eccentricities clustered around $\sim 0.1$ (e.g., PSR~J2234+0611, \citealt{deneva2013goals}; PSR~J0955$-$6150, \citealt{camilo2015parkes}), and a plausible explanation is that their formation may be linked to ejected material from the companion (e.g., \citealt{antoniadis2014formation}). A rarer evolutionary channel involves the disruption of triple systems, as in the case of PSR~J1903+0327, whose eccentricity $\sim 0.44$ is thought to result from the ejection of its inner companion (\citealt{freire2011nature}; \citealt{zwart2011formation}). 

In recent years, observations with the Five-hundred-meter Aperture Spherical radio Telescope (FAST; \citealt{nan2011five}) have greatly expanded the pulsar sample, leading to the discovery or confirmation of more than 60 new pulsars in GCs\footnote{\url{https://fast.bao.ac.cn/cms/article/65/}} (\citealt{lian2025fast}), including the first two MSPs identified in M10 (NGC 6254; \citealt{pan2021fast}). M10 is a metal-poor GC with R.A. $\alpha_{\rm J2000}$ = $16^{{\rm h}} 57^{{\rm m}} 09^{{\rm s}}.05$ and decl. $\delta_{\rm J2000} = -04^\circ 06\arcmin 01\arcsec.1$ (\citealt{har96, harris2010new}), at a heliocentric distance of $\sim$ 5.1 kpc (\citealt{baumgardt2021accurate}). It has a core radius $r_c = 0.81$ pc ($0.55\arcmin$) and an estimated density $\log \rho_c [{\rm M}_\odot/{\rm pc}^3] =$ 3.78 (\citealt{baumgardt2018catalogue}), which represents a moderately high central density compared to other GCs (where $\log \rho_c$ spans from $\sim -1$ to $\sim 6$).

PSR~J1657$-$0406A (M10A) is the first pulsar discovered in M10. It has a spin period of 4.73 ms, a dispersion measure (DM) of 43.9 pc cm$^{-3}$, and exhibits scintillation phenomena that were unconfirmed at the time (\citealt{pan2021fast}). Its observed period variations strongly suggest that M10A is in a binary system. 

In this paper, we present the timing solution of M10A based on nearly 7 yr of observations taken with FAST. Section \ref{sec:observations} shows the observational setup and data reduction procedures. Section \ref{sec:results} presents the timing analysis and mass inferences, while Section \ref{sec:discussion} discusses the possible evolutionary history and the nature of the companion. Conclusions are summarized in Section \ref{sec:conclusions}.

\section{Observations and Data Reduction}
\label{sec:observations}

M10 was observed with FAST from 2018 December to 2025 July, for a total of 35 observation sessions. The data analyzed in this work come from different projects, thus resulting in nonuniform intervals between consecutive sessions (from 1 to 700 days), and integration times (from 1.0 to 3.5 hr). The observations were conducted using the 19-beam \textit{L}-band receiver, but we considered the data taken with the central beam only, since it has a size of $\sim 3\arcmin$ and effectively covers the core region of M10. The signal in the frequency range from 1.0 to 1.5 GHz is divided into 4096 channels with a width of 0.122 MHz (\citealt{jiang2020fundamental}). Data were recorded using dual-polarization 8-bit sampling, with a sampling time of 49.152 $\mu$s.

We processed the data using the \texttt{PRESTO} software package\footnote{\url{https://github.com/scottransom/presto}} (\citealt{2011ascl.soft07017R}) and dedispersed them using the approximate DM values from \citet{pan2021fast}, allowing the pulsar signal to be searched for and folded. We created a standard noise-free pulse template by identifying the average pulse profile with the highest signal-to-noise ratio (SNR) and by fitting it against a model consisting of an appropriate number of Gaussian components, using the \texttt{pygaussfit.py} routine. We then obtained the pulses' times of arrival (ToAs) by cross-correlating in the Fourier domain the observed pulses' profiles with this template using the \texttt{get\_TOAs.py} routine (for the method, see \citealt{taylor1992pulsar}). Our ToA extraction strategy followed an empirical approach. We considered both the orbital coverage on one side, aiming to get more ToAs per observation for a better description of the orbit, and the ToA errors on the other side, in order to have similar errors from all observations, which we kept around 50 $\mu$s, i.e. about $1\%$ of the pulsar period. As a result, we extracted more ToAs from observations with higher SNRs or longer integration times and fewer from those with poorer conditions. 

We conducted the timing analysis using \texttt{TEMPO} (\citealt{2015ascl.soft09002N}). We derived a preliminary orbital solution from the observed variations of the pulsar spin period, using the \texttt{fitorb.py} routine. Then, we iteratively optimized the pulsar ephemeris through ToA fitting. We employed the DE438 solar system ephemeris model\footnote{\url{https://naif.jpl.nasa.gov/pub/naif/JUNO/kernels/spk/de438s.bsp.lbl}}, UTC (NIST) clock standard and Barycentric Dynamical Time (TDB) units. We described the orbital motion using the binary model DD (\citealt{1985AIHPA..43..107D}, \citealt{1986AIHPA..44..263D}) , which allows the measurement of the Keplerian parameters, and also of the post-Keplerian (PK) ones including the periastron advance ($\dot\omega$). We employed the \texttt{Dracula}\footnote{\url{https://github.com/pfreire163/Dracula}} script (\citealt{freire2018algorithm}) to construct phase-connected solutions, including precise position, spin, and orbital parameters. During this process, the DM of the pulsar was also optimized by fitting ToAs obtained from segmented frequency bands.

\section{Timing and Mass Measurements} \label{sec:results}

\begin{table*}[!htbp]
\centering
\caption{Timing Solution of M10A}
\label{timing_parameters}
\begin{tabular}{cc}
\hline\hline
Parameter               &      Value         \\
\hline
\multicolumn{2}{c}{Timing Parameters} \\
\hline
Right ascension, R.A. (J2000)               \dotfill   &  16:57:10.2720(2)        \\
Decl., decl. (J2000)                        \dotfill   &  $-$04:05:59.858(4)         \\
Proper motion in R.A., $\mu_\alpha \cos{\delta}$ (mas yr$^{-1}$)            \dotfill   &    $-$1.8(17)       \\
Proper motion in decl., $\mu_\delta$ (mas yr$^{-1}$)                        \dotfill   &    $-$6(3)      \\
Parallax$^{a}$, $\varpi$ (mas)              \dotfill   &    0.1960      \\
Spin frequency, $f$ (Hz)                    \dotfill   &  211.23477609966(2)         \\
First spin frequency derivative, 
$\dot{f} (10^{-16} {\rm s}^{-2})$           \dotfill   &  1.542(5)               \\
Dispersion measure, DM (cm$^{-3}$~pc)       \dotfill   &  43.885(4)                                    \\
Binary model                                \dotfill   &  DD                             \\
Orbital period, $P_{\rm b}$ (days)          \dotfill   &  37.3986270(14)                \\
Projected semi-major axis, $x$ (lt-s)       \dotfill   &  14.246407(3)                   \\
Epoch of periastron, $T_0$ (MJD)            \dotfill   &  59859.56355(2)                  \\
Orbital eccentricity, $e$                   \dotfill   &  0.1192287(4)                \\
Longitude of periastron, $\omega$ (deg)     \dotfill   &  245.2969(2)               \\
Periastron advance rate, 
$\dot{\omega}$ (deg  yr$^{-1}$)             \dotfill   &  0.00090(13)                \\
Solar System ephemeris                      \dotfill   &  DE438                      \\
Reference epoch (MJD)                       \dotfill   &  60000                     \\
Start of timing data (MJD)                  \dotfill   &  58477.119                               \\
End of timing data (MJD)                    \dotfill   &  60864.591                           \\
Number of ToAs                              \dotfill   &  563                    \\
Weighted rms timing residuals ($\mu$s)      \dotfill   &  34.81             \\
\hline 
\multicolumn{2}{c}{Derived Parameters} \\
\hline
Spin period, $P$ (ms)                                              \dotfill &  4.7340689751208(4)        \\
First spin period derivative, $\dot{P}$ ($10^{-21}$ s s$^{-1}$)    \dotfill &  $-$3.456(11)                  \\ 
Mass function, $f_{\rm M}$ (${\rm M}_\odot$)                       \dotfill &  0.0022196651(14)                 \\
Total system mass (assuming GR), $M_{\rm tot}$ (${\rm M}_\odot$)   \dotfill &  2.5(6)                   \\
Maximum pulsar mass$^{b}$, $M_{\rm p,max}$ (${\rm M}_\odot$)       \dotfill &  2.9                      \\
Minimum companion mass$^{b}$, $M_{\rm c,min}$ (${\rm M}_\odot$)    \dotfill &  0.20                 \\
Orbital separation, $a$ (lt-s)                                     \dotfill &  149(11)                        \\
\hline
\end{tabular}
\begin{center}
{\footnotesize \textbf{Notes.} Values in parentheses represent the 1$\sigma$ uncertainties in the last digits quoted, as determined by the \textsc{TEMPO} fitting procedure.}\\ 
{\footnotesize $^{a}$ Fixed value from \citet{vasiliev2021gaia};}\\
{\footnotesize $^{b}$ Combining $f_{\rm M}$ with $M_{\rm tot}$ and assuming $i=90^\circ$.}
\end{center}
\label{tab:solution}
\end{table*}

\begin{figure}[!htbp]
\centering
\epsscale{1.2}
\plotone{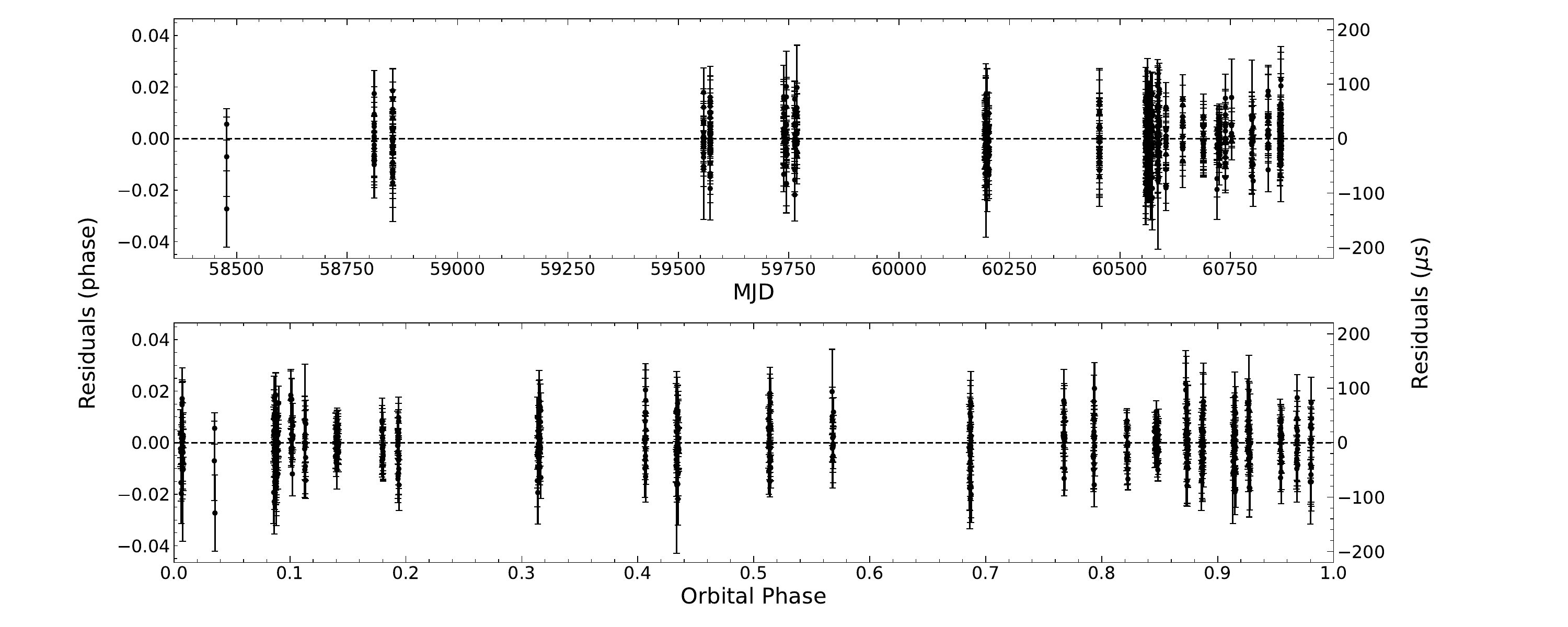}
\caption{Postfit timing residuals vs. MJD (top) and orbital phase (bottom).}
\label{fig:Res}
\end{figure}

Our timing analysis confirms that M10A is a member of a long period ($P_{\rm b} \sim$ 37 days), wide ($x \sim$ 14 lt-s), and eccentric ($e \sim$ 0.12) binary system. Using these orbital parameters, we derived the mass function $f_{\rm M} = (M_{\rm c} \sin i)^3/(M_{\rm p} + M_{\rm c})^2 = 4 \pi^2 x^3/(P_{\rm b}^2 T_\odot ) \ {\rm M}_\odot = 0.0022196651(14) \ {\rm M}_\odot$ (here and throughout the paper, numbers in parentheses represent the $1\sigma$ uncertainty in the last significant digit), where $T_\odot = G \ {\rm M}_\odot/c^3 = 4.925490947 \ \mu {\rm s}$, $i$ is the orbital inclination, and $M_{\rm p}$ and $M_{\rm c}$ represent the masses of the pulsar and its companion, respectively. These parameters were obtained without strong constraints on other kinematic parameters (proper motion and parallax) or higher-order spin derivatives (particularly the second frequency derivative) based on current timing precision. Nevertheless, the derived mass function is insensitive to these unconstrained parameters and is therefore robust at the quoted precision.

\begin{figure}[!h]
\centering
\epsscale{0.75}
\plotone{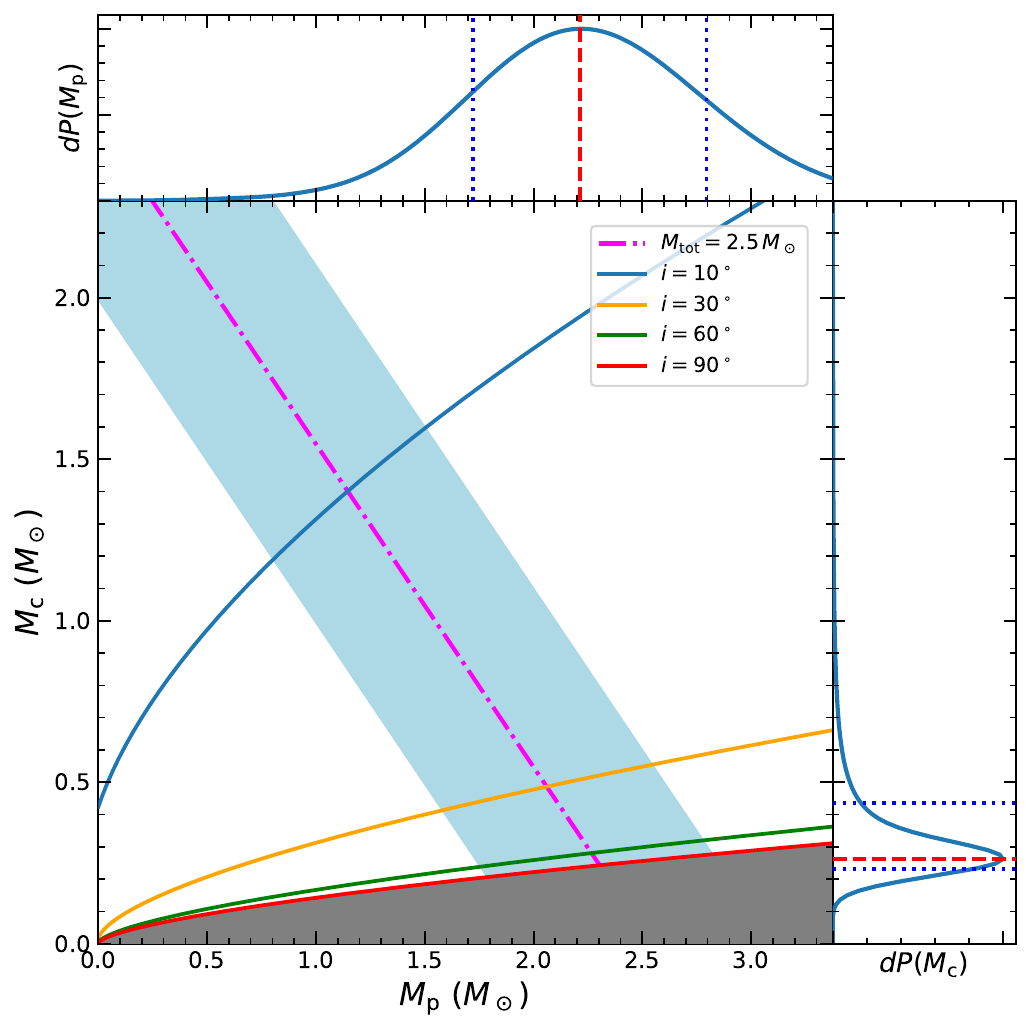}
\caption{Mass-mass diagram for M10A. Main panel: $M_{\rm p}$--$M_{\rm c}$ space; the magenta dot-dashed line indicates the $\dot{\omega}$-derived total mass, with its $1\sigma$ uncertainty shaded in blue; the gray region is excluded by the condition $\sin i \le 0$; the $M_{\rm p}$--$M_{\rm c}$ relations are shown for inclinations $i = 10^\circ$, $30^\circ$, $60^\circ$, and $90^\circ$ in blue, orange, green, and red, respectively. Top and right panels: $M_{\rm p}$ and $M_{\rm c}$ 1D PDFs, respectively; red dashed lines mark the most probable values, while blue dotted lines indicate the 16th/84th percentiles ($1\sigma$ uncertainties). Units for the 1D PDFs are arbitrary.}
\label{fig:pdf}
\end{figure}

We also attempted to measure the PK parameters of this binary system, in order to obtain constraints on the mass of both the pulsar and the companion. Owing to the relatively large orbital eccentricity, we measured the periastron advance rate $\dot{\omega} = 0.00101(11)$ deg yr$^{-1}$. Assuming the observed periastron advance is entirely due to general relativity (GR), the total mass is obtained from the following equation:
\begin{equation}
\frac{M_{\rm tot}}{M_\odot} = \frac{1}{T_\odot} \left[ \frac{\dot{\omega}}{3}(1 - e^2) \right]^{3/2} \left( \frac{P_{\rm b}}{2\pi} \right)^{5/2} \label{eq:omega_dot}
\end{equation}
This yields $M_{{\rm tot}} = 3.0(5)\,{\rm M}_\odot$. Among the currently measured total masses of binary pulsar systems\footnote{\url{https://www3.mpifr-bonn.mpg.de/staff/pfreire/NS_masses.html}}, it is exceeded only by PSR~J0514$-$4002E in NGC 1851 ($M_{{\rm tot}} = 3.8870(45)\ {\rm M}_\odot$; \citealt{barr2024pulsar}), PSR~J1748$-$2446ao in Terzan 5 ($M_{{\rm tot}} = 3.166(24)\ {\rm M}_\odot$; \citealt{padmanabh2024discovery}), and PSR~J2222$-$0137 ($M_{{\rm tot}} = 3.150(14)\ {\rm M}_\odot$; \citealt{kaplan20141}). However, the mass of this system may be overestimated. Although $\dot\omega$ is measured with a 10$\sigma$ significance, its value may still be contaminated by unmodeled effects. We addressed these possible contaminations by repeating our timing analysis using different constraints on the proper motion and the parallax. We thus inserted in our timing model the values for the proper motion in R.A. ($\mu_\alpha \cos\delta$) and decl. ($\mu_\delta$), and parallax ($\varpi$). At first we used the values for the host GC, namely $\mu_\alpha \cos{\delta} = -4.758$ mas yr$^{-1}$, $\mu_\delta = -6.597$ mas yr$^{-1}$ and $\varpi = 0.196$ mas \citep{vasiliev2021gaia}, and we kept them fixed while fitting for all other timing parameters. We thus obtained $\dot{\omega} = 0.00079(11)$ deg yr$^{-1}$ and $M_{{\rm tot}} = 2.1(4) \ {\rm M}_\odot$. The total mass is now lower by about the $30 \%$, but remains questionable since the measured proper motions of pulsars may differ from those of their hosts, as the cases in M2 \citep{li2025millisecond} and Terzan 5 \citep{padmanabh2024discovery}. Therefore, we decided to keep fixed the parallax to the GC value, since the pulsar distance to the GC center is negligible compared to the system distance, and fit for the proper motion components. We thus obtained $\dot{\omega} = 0.00090(13)$ deg yr$^{-1}$ and $M_{{\rm tot}} = 2.5(6) \ {\rm M}_\odot$. The fitted proper motions are $\mu_\alpha \cos{\delta} = -1.8(17)$ mas yr$^{-1}$ and $\mu_\delta = -6(3)$ mas yr$^{-1}$, with the latter being better constrained. The corresponding full timing solution is presented in Table~\ref{tab:solution}, and the related postfit timing residuals are displayed in Figure~\ref{fig:Res}. Because of the motivations mentioned above, we considered this last timing solution as the physically most reasonable one, and we adopted it as a reference for our subsequent analysis. We also attempted to measure $\ddot{f}$, but we found values with uncertainties larger than the value itself, thus suggesting unreliable detections. Combining $f_{\rm M}$ and $M_{\rm tot}$ and assuming $i=90^\circ$, we obtained $M_{\rm p,max} = 2.9 \ {\rm M}_\odot$ and $M_{\rm c,min} = 0.20 \ {\rm M}_\odot$ at the $1\sigma$ confidence level. We also applied the Bayesian methods described in \citet{splaver2002probing} and \citet{freire2011nature}, using the $\chi^2$ from the TEMPO timing fit, to derive the posterior probability distributions of the component masses. The resulting values are $M_{{\rm p}} = 2.2^{+0.6}_{-0.5}\ {\rm M}_\odot$ and $M_{{\rm c}} = 0.26^{+0.17}_{-0.03}\ {\rm M}_\odot$ (1$\sigma$ uncertainties, calculated as the 16th/84th percentiles), indicating a massive NS with a HeWD companion. Figure~\ref{fig:pdf} shows the $M_{\rm p}$–$M_{\rm c}$ plane, along with the probability density functions (PDFs) of $M_{\rm p}$ and $M_{\rm c}$. The posterior distribution indicates a $\sim$ 67\% probability that the pulsar mass exceeds $2.0 \ {\rm M}_\odot$, making M10A a candidate for one of the most massive NSs. The estimated mass of M10A is comparable to that of PSR~J0740+6620 ($2.08(7) \ {\rm M}_\odot$; \citealt{fonseca2021refined}), which remains among the most massive pulsars with a mass measured via the Shapiro delay. Alternatively, if M10A follows the typical MSP mass distribution (e.g., \citealt{antoniadis2016millisecond}), the system would likely be a nearly face-on binary, either a massive WD–NS system or a double NS (DNS) system. Adopting the lowest precisely measured NS mass, $1.174(4) \ {\rm M}_\odot$ \citep{martinez2015pulsar}, $i$ would be equal to or larger than 10$^\circ$. In these cases, a secondary exchange would have occurred after the speeding up of the pulsar into an MSP. This scenario appears more physically plausible. However, a massive companion may struggle to account for the eccentricity (see Section~\ref{sec:exchange}).

\section{Discussion} \label{sec:discussion}

Mass accretion during MSP formation generally circularizes the orbit. The notable eccentricity ($e \sim 0.12$) of M10A, however, is unusual and deserves further investigation. In this section, we investigated the nature of M10A companion, the origin of the orbital eccentricity, and possible dynamical formation mechanisms.

\subsection{Searches for an optical counterpart}  \label{sec:Optical}

In order to shed light on the nature of the companion, we searched for an optical counterpart of M10A in optical archival data. We analyzed at first Hubble Space Telescope (HST) images of the M10A field. Observations were obtained with the Wide Field Camera 3 (WFC3; \citealt{kimble2008wide}) using the F275W, F336W, and F438W filters (GO13297, PI Piotto), and with the Advanced Camera for Surveys (ACS; \citealt{ford1998advanced}) using the F606W and F814W filters (GO10775, PI: Sarajedini; SNAP16298, PI: Libralato for F814W). We also queried the Mikulski Archive for Space Telescopes (MAST)\footnote{\url{https://mast.stsci.edu/search/ui/\#/}} for calibrated data products. In all of these attempts, no significant point sources were detected at the pulsar position.

\begin{figure}[!htpb]
\centering
\epsscale{1.2}
\plotone{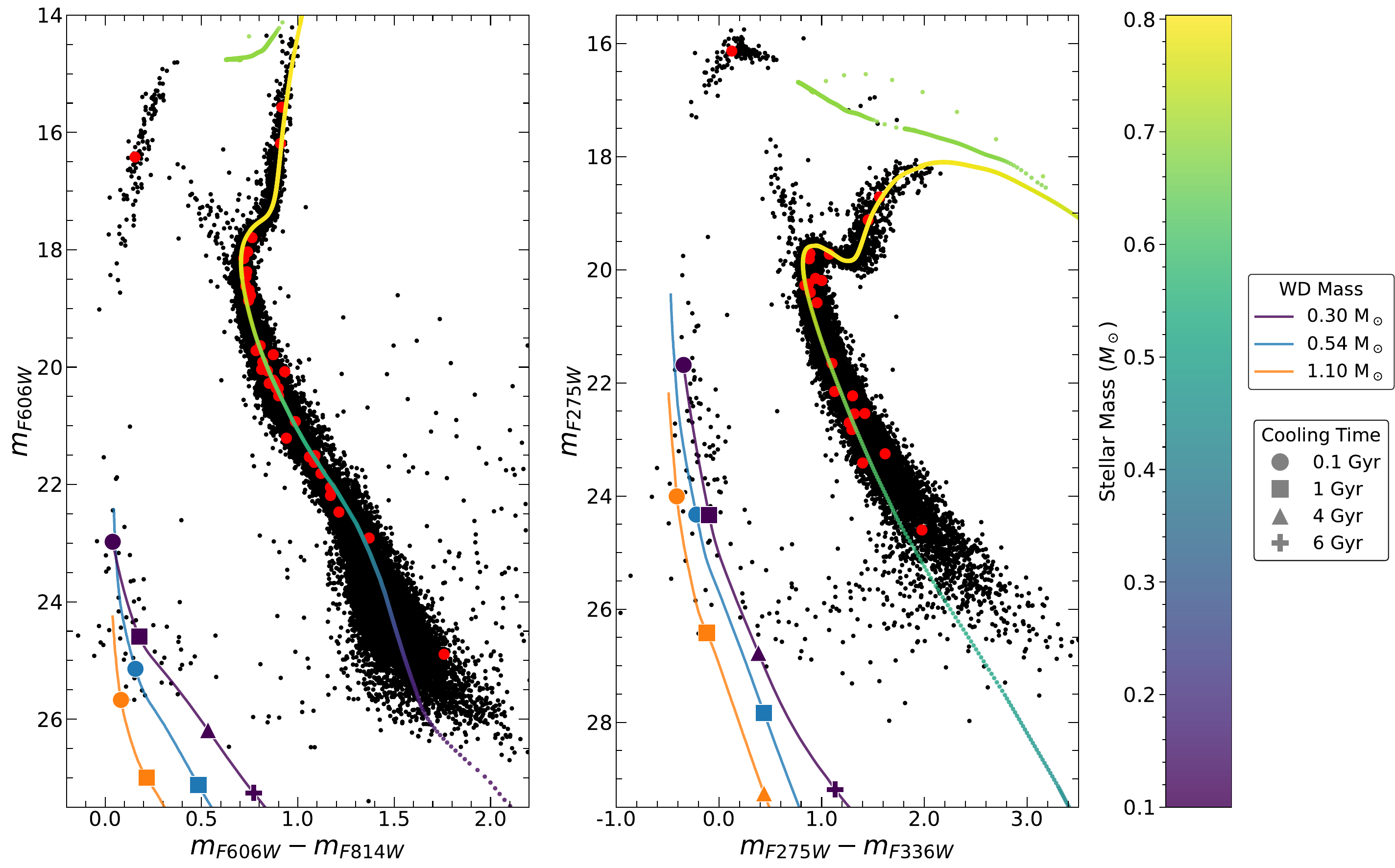}
\caption{Color$-$magnitude diagrams (CMDs) of M10 from HUGS Survey: $m_{\rm F606W}$ vs. $m_{\rm F606W} - m_{\rm F814W}$ (left) and $m_{\rm F275W}$ vs. $m_{\rm F275W} - m_{\rm F336W}$ (right). Red points mark sources within a $5\arcsec \times 5\arcsec$ region centered on M10A. The gradient-colored line represents a 12.3 Gyr isochrone, with its color varying according to the stellar mass (reference scale displayed in the rightmost panel). The other curves represent the cooling tracks for a 0.3 ${\rm M}_\odot$ HeWD (purple) as well as 0.54 ${\rm M}_\odot$ (blue) and 1.10 ${\rm M}_\odot$ (orange) COWDs.}
\label{fig:CMD}
\end{figure}

Given our non-detection of any optical counterpart, we performed a deeper photometric analysis of the stars in M10. We thus constructed color-magnitude diagrams (CMDs) of M10 using the HST UV Globular Cluster Survey catalog data (HUGS\footnote{\url{https://archive.stsci.edu/prepds/hugs}}; \citealt{piotto2015hubble}; \citealt{nardiello2018hubble}; \citealt{hugs}, programs GO10775 and GO13297), which provides photometric data in the above five filters. To select reliable sources, we restricted the samples within the half-light radius ($2.03 \arcmin$; \citealt{baumgardt2018catalogue}) with photometric RMS $<1.5$ mag. We also imposed the image shape parameter (RADXS; \citealt{bedin2008reaching}) to be $|$RADXS$|<0.15$, and further selected only those sources with membership probability either $\geq 90\%$ or not available \citep{nardiello2018hubble, pichardo2025second}, to ensure clean point-source detections. Figure \ref{fig:CMD} shows the CMDs of $m_{{\rm F606W}}$ versus $m_{{\rm F606W}}-m_{{\rm F814W}}$ and $m_{{\rm F275W}}$ versus $m_{{\rm F275W}}-m_{{\rm F336W}}$. Red points mark sources within $5\arcsec \times 5\arcsec$ of M10A, including main sequence (MS), subgiant, and horizontal branch (HB) stars, but we could not identify any reliable WD candidate. 

In this case, we need to consider the possibility that the companion may be too faint for detection. One possibility is that it is a low-mass MS star, in which case stellar isochrones are required to convert the HST photometric limits into constraints on its mass. We adopted theoretical isochrones from the BaSTI database\footnote{\url{http://basti-iac.oa-abruzzo.inaf.it/index.html}} \citep{pietrinferni2006large, pietrinferni2021updated} with metallicity [Fe/H] = $-$1.56 (\citealt{har96, harris2010new}). Taking the 11.39 Gyr age from \citet{forbes2010accreted} as a reference, we adopted a slightly older 12.3 Gyr isochrone to account for the omission of atomic diffusion in the BaSTI models, which can increase the inferred ages by 0.8–0.9 Gyr (e.g., \citealt{kerber2018ages, oliveira2020hubble}). For comparison with the HUGS photometry, we adopted reddening \textit{E}(\textit{B}$-$\textit{V}) = 0.28 from \citet{har96, harris2010new} and filter-specific extinction coefficients from \citet{cardelli1989relationship} and \citet{o1994rnu}. By visual inspection, we determined that a distance modulus of $(m-M)_0 = 13.5$ provides a satisfactory benchmark for the CMDs, consistent with $\sim 13.5$ from \citet{baumgardt2021accurate}. The comparison indicates that MS stars above $\sim 0.4\ {\rm M}_\odot$ are detectable in both V and near-UV; therefore, a companion with a minimum mass of $M_{\rm c,min} = 0.20\ {\rm M}_\odot$ is expected to be marginally observable in HST optical bands. However, if the companion is a low-mass MS star, the binary orbit would likely have a high inclination. In the absence of radio eclipses, this scenario is disfavored.

Considering the HST observations only, M10A companion could be either an NS or a cool WD. To further assess this possibility, we followed the approach of \citet{cadelano2019extremely, cadelano2020psr}. We adopted the 0.54 and 1.10 ${\rm M}_\odot$ COWD cooling tracks from BaSTI \citep{salaris2010large, salaris2022updated}, adopting electron conduction opacities from \citet{blouin2020new} and assuming a hydrogen atmosphere. The tracks were placed on the CMDs using the same procedure as for the isochrone. The comparison shows that if the WD cooling time is larger than 1 Gyr, the WD would be too faint to be detected in the HST data. For HeWDs, we adopted the 0.302 ${\rm M}_\odot$ cooling track of \citet{istrate2014timescale} and \citet{istrate2016models}, combined with the Koester WD spectra templates \citep{tremblay2009spectroscopic, koester2010white}, and we used the Python package \texttt{PySynphot}\footnote{\url{https://pysynphot.readthedocs.io/en/latest/}} \citep{2013ascl.soft03023S} to transform theoretical luminosities into HST magnitudes in various filters. By setting the color index and distance modulus to the values for M10, we conservatively estimated that a $\sim 0.3\ {\rm M}_\odot$ HeWD in M10 would require $\sim 6$ Gyr of cooling time, which is well within the lifetime of M10 \citep[11.39 Gyr,][]{forbes2010accreted}.

Finally we examined archival data from the Chandra Observatory and the Fermi Large Area Telescope (LAT) catalogs, but found no X-ray or $\gamma$-ray source consistent with the pulsar timing position. The lack of strong high-energy emission suggests that there is no ongoing accretion or strong shock emission, consistent with a late-stage binary system hosting an old pulsar with a degenerate companion \citep[e.g.,][]{caraveo2014gamma, zhao2022census}.

\subsection{Dynamical Exchange} \label{sec:exchange}

M10A may have undergone a secondary exchange encounter, in which its original low-mass companion first recycled the pulsar into an MSP and was subsequently replaced by the current compact companion, producing the present eccentric binary. Such exchange encounters are thought to explain the high incidence of eccentric binaries in GCs, particularly those with high encounter rates (\citealt{verbunt2014disruption}, e.g., PSR~J0514$-$4002A, $e \sim 0.89$, \citealt{ridolfi2019upgraded}). However, dynamical encounters tend to produce higher eccentricities. Currently, the lowest-eccentricity MSP system believed to have formed via a secondary exchange is Terzan 5ao with $e \sim 0.32$ (\citealt{padmanabh2024discovery}), still far above that of M10A. Therefore, whether M10A originated from a secondary exchange remains uncertain.

Following the approach of \citet{decesar2015highly}, we used the FEWBODY\footnote{\url{https://github.com/YihanWangAstro/Fregeau-Fewbody}} \textit{N}-body code \citep{fregeau2004stellar, 2012ascl.soft08011F} to model encounters between a naturally evolved circular MSP$-$HeWD binary and a third body, that we assumed to be either a $1.3 \ {\rm M}_\odot$ NS, a $1.2 \ {\rm M}_\odot$ ONeMgWD, a $0.7 \ {\rm M}_\odot$ COWD, and a $0.3 \ {\rm M}_\odot$ HeWD. For each possible third body nature, we performed 5000 simulations, drawing the encounter velocities from a Maxwellian distribution (velocity dispersion $\sigma_0=6.2$ km s$^{-1}$ for M10; \citealt{baumgardt2018catalogue}) and impact parameters up to 30 times the initial orbital separation. The initial $P_{\rm b}$ and the corresponding HeWD companion mass were constrained by using the $P_\mathrm{b}$–$M_\mathrm{WD}$ correlation from \citet[eqs. 20--21]{tauris1999formation}. To ensure consistency with M10A, we constrained the simulated postencounter $P_{\rm b}$ to have the peak of the kernel density estimate (KDE) of its distribution within $\pm 5\%$ of the $P_{\rm b}$ of M10A, and derived initial $P_{\rm b}$ values of 4, 5, 10, and 42 days for the four types of third bodies, respectively. Exchange probabilities are 66\%, 68\%, 60\%, and 16\% for the four types of third bodies, respectively, but postexchange systems rarely are nearly circular: low-eccentricity outcomes ($e<0.15$) occur only in the 0.06\%, 0.15\%, 1.0\%, and 2.4\% of cases. Therefore, exchange encounters rarely produce low-eccentricity binaries, but such a scenario cannot be completely ruled out in the case of M10A.

The encounter rates governing MSP dynamics scale as $\Gamma \propto \rho_{\rm c}^{1.5} r_{\rm c}^2$ (two-body) and $\gamma \propto \rho_{\rm c}^{0.5} r_{\rm c}^{-1}$ (single–binary; \citealt{verbunt1987globular, verbunt2002binary}). A higher $\gamma$ increases the probability of exchange encounters, promoting eccentric binaries and binary disruption that can yield isolated MSPs. The reported $\gamma$ is generally normalized to the value for M4 (\citealt{verbunt2014disruption}). Using the parameters from \citet{harris2010new}, we obtained for M10 $\gamma_{\rm M10} \sim 0.7\,\gamma_{\rm M4}$. Following the method of \citet{decesar2015highly}, we estimate the binary–single encounter rate as $\xi_{1+2,{\rm M10}} \sim 0.1\ {\rm Gyr}^{-1}$, indicating a relatively quiescent dynamical environment. By contrast, NGC 1851 has $\gamma_{\rm NGC 1851} \sim 12.4\,\gamma_{\rm M4}$ and $\xi_{1+2,{\rm NGC 1851}} \sim 2\ {\rm Gyr}^{-1}$, consistent with the presence of three highly eccentric binary MSPs (A, D, E; \citealt{fregeau2004stellar, ridolfi2022trapum, barr2024pulsar}).

\subsection{A WD Companion and the Origin of the Eccentricity} \label{sec:Eccentricity}

It is important to consider the scenario in which M10A has not undergone any strong dynamical encounter since the end of the accretion phase. In this case, the present companion would be the star that recycled the pulsar. An NS companion can be ruled out. Although a supernova kick can produce an eccentric orbit, mass transfer in high-mass X-ray binaries (HMXBs) is typically brief and inefficient \citep{belczynski2010double, tauris2017formation, chattopadhyay2020modelling, deng2024formation}, resulting in only mildly recycled pulsars with spin periods in the range of 20–100 ms (e.g., PSR~J1411+2551, $P \sim 62$ ms, \citealt{martinez2017pulsar}; PSR~J1829+2456, $P \sim 41$ ms, \citealt{champion2004psr}). Intermediate-mass X-ray binaries (IMXBs), which produce CO or ONeMgWD companions, generally produce only partially recycled pulsars \citep{tauris2011five, tauris2012formation}, although a few low-mass ($< 0.5\ {\rm M}_\odot$) COWDs are known to host MSPs with spin periods $P<10$ ms (e.g., PSR~J1125$–$6014, \citealt{shamohammadi2023searches}; PSR~J1933$–$6211, \citealt{geyer2023mass}). By contrast, the vast majority of MSPs are thought to originate from low-mass X-ray binaries (LMXBs; \citealt{tauris2012formation}). If the companion is an NS or WD with $M > 0.6 \ {\rm M}_\odot$, the mass function implies a nearly face-on orbit ($i < 24^\circ$). Such an orientation has a relatively low prior probability, not larger than the 9\%. Therefore, if the present companion is indeed the star that recycled the pulsar, it is most likely either a low-mass COWD or a HeWD. In both cases, the mass transfer process is expected to efficiently circularize the orbit through tidal interactions and viscous dissipation in the envelope \citep{verbunt1993origin, phinney1994binary}. The residual eccentricities are typically tiny ($e \sim 10^{-6}$–$10^{-3}$; \citealt{phinney1992pulsars}), much smaller than $e \sim 0.12$ of M10A. Thus, the significant eccentricity most likely reflects dynamical processes occurring after the recycling phase.

\subsubsection{Gravitational Perturbation} \label{sec:perturbation}

In GCs, repeated encounters with passing stars can gradually increase the orbital eccentricities of wide low-mass binary MSPs. Even the highly eccentric system PSR~J1748$-$2446ap in Terzan 5 ($e \sim 0.91$) has been suggested to originate through this process \citep{padmanabh2024discovery}. The eccentricity growth timescale can be estimated as \citep{rasio1995orbital}:
\begin{equation}
t_{\rm e} \simeq 2 \times 10^{11} {\rm yr} \left(\frac{n}{10^4 \ {\rm pc}^{-3}}\right)^{-1}\left(\frac{\sigma_0}{10 \ {\rm  km \ s}^{-1}}\right) \left(\frac{P_{\rm b}}{{\rm days}}\right)^{-2/3} [\ln (4/e)]^{-2/3},
\label{eq:ecc_tub}
\end{equation}
where $n$ is the stellar number density, $\sigma_0$ is the central one-dimensional velocity dispersion, $P_{\rm b}$ is the orbital period, and $e$ is the orbital eccentricity. We also assumed $n \propto \rho_{\rm c}$ as in, e.g., \citet{lian2023discovery}. We adopted NGC 6517 as a reference cluster, for which $\rho_{\rm c} \sim 1.95 \times 10^5 \ {\rm L}_{\odot} \ {\rm pc}^{-3}$ and $n \sim 4.7 \times 10^5 \ {\rm pc}^{-3}$ \citep{lynch2011six}. Using for M10 $\rho_{\rm c} \sim 3.47 \times 10^3 \ {\rm L}_{\odot} \ {\rm pc}^{-3}$ \citep{harris2010new}, we obtain $n \sim 8.36 \times 10^3 \ {\rm pc}^{-3}$. Adopting $\sigma_0=6.2$ km s$^{-1}$ \citep{baumgardt2018catalogue}, the resulting eccentricity growth timescale for M10A is $t_{\rm e} \sim 5.7$ Gyr. Given the cluster age of 11.39 Gyr \citep{forbes2010accreted}, M10A has had sufficient time for gravitational perturbations to raise its orbital eccentricity.

\subsubsection{Does M10A Belong to a Peculiar Eccentric MSP$–$HeWD Population?} \label{sec:eccMSP-HeWD}

\begin{figure}[htpb]
\centering
\epsscale{0.88}
\plotone{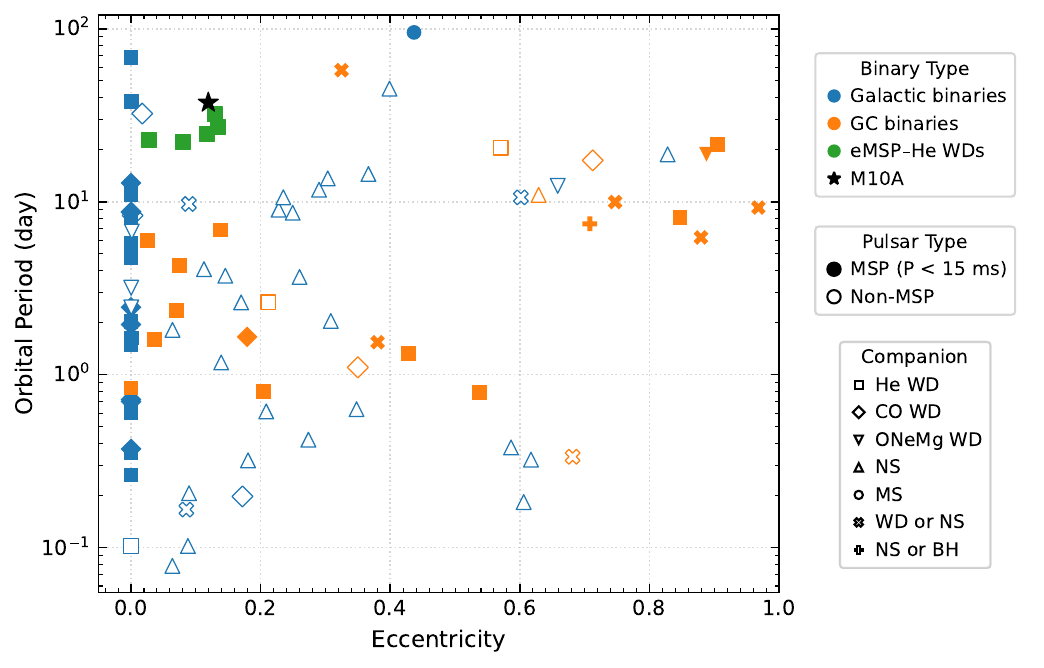}
\caption{Orbital period vs. eccentricity of pulsar binaries with no mass transfer or significant mass loss. The position of M10A is indicated by the black star. Green points denote the eMSP$-$HeWDs binaries summarized in Table 2. Orange points represent systems located in GCs, while blue points denote those in the Galactic field. Filled and unfilled symbols are for MSP and non-MSP systems, respectively, with different shapes indicating companion types. \\
\textbf{References:} https://www3.mpifr-bonn.mpg.de/staff/pfreire/NS\_masses.html and references therein.
\label{fig:eccMSP}
}
\end{figure}

As discussed above, M10A may host a HeWD companion formed through the evolution of its donor star. Over the past decade, several eccentric MSP$–$HeWD systems (eMSP$–$HeWDs) have been discovered in the Galactic field. These systems typically occupy a narrow range of orbital parameters ($e \sim 0.02$–$0.13$, $x \sim 10$–$14$ lt-s, $P_{\rm b} \sim 22$–$32$ days), but pulsar masses span a broader range. M10A is very similar to these systems, as shown in Table \ref{tab:eccentric_MSPs}, where its parameters are reported along with the ones for these systems in the Galactic field. The similarity of M10A with these peculiar systems is also evident in Figure \ref{fig:eccMSP}, where the orbital eccentricity is plotted against the orbital period for recycled binary pulsars with no mass transfer or significant mass loss. M10A (black star) is located very close to the known eMSP$–$HeWDs in the Galactic field (green points), showing similar $e-P_{\rm b}$ combinations, while being separated from other GC MSPs (orange filled). This suggests that M10A may belong to the peculiar eMSP$-$HeWD population, and that its eccentricity may originate from processes internal to the binary rather than from dynamical interactions within the GC.

\begin{table}[htpb]
\centering
\caption{Orbital and mass parameters of Known eMSP$–$HeWD systems and M10A}
\begin{adjustbox}{center=\textwidth,right=16cm}
\setlength{\tabcolsep}{3mm}{
\begin{tabular}{ccccccccc}
\hline\hline
Pulsar & $P$  & $P_{\rm b}$  & $x$  & $e$ & $M_{\rm tot}$  & $M_{\rm p}$ & $M_{\rm c}$  & References \\
 &  (ms) &  (days) &  (lt-s) &  &  (${\rm M}_\odot$) &  (${\rm M}_\odot$) & (${\rm M}_\odot$) & \\
\hline
J0955$-$6150  & 2.00  & 24.6 & 13.3 & 0.12 & 1.96(3) & 1.71(2) & 0.254(2) & (1, 2) \\
J1618$-$3921  & 11.99 & 22.7 & 10.3 & 0.03 & 1.42$^{+0.20}_{-0.19}$ & 1.20$^{+0.19}_{-0.20}$ & 0.20$^{+0.11}_{-0.03}$ & (3-5) \\
J1657$-$0406A & 4.73  & 37.4 & 14.2 & 0.12 & 2.5(6) & $<$ 2.9 & $>$ 0.20 & this work \\
J1946+3417    & 3.17  & 27.0 & 13.9 & 0.13 & 2.094(22) & 1.828(22) & 0.2656(20) & (6, 7) \\
J1950+2414    & 4.30  & 22.2 & 14.2 & 0.08 & 1.779(25) & 1.496(23) & 0.2795$^{+0.0046}_{-0.0038}$ & (8-10) \\
J2234+0611    & 3.58  & 32.0 & 13.9 & 0.13 & 1.6518$^{+0.0033}_{-0.0035}$ & 1.353$^{+0.014}_{-0.017}$ & 0.298$^{+0.015}_{-0.012}$ & (11, 12) \\
\hline
\end{tabular}}
\end{adjustbox}
\label{tab:eccentric_MSPs}
\small
\begin{itemize} 
\item References. (1) \citet{camilo2015parkes}; (2) \citet{serylak2022eccentric}; 
(3) \citet{edwards2001recycled}; (4) \citet{octau2018psr}; (5) \citet{grunthal2024triple};
(6) \citet{barr2013northern}; (7) \citet{barr2017massive}; 
(8) \citet{allen2013einstein}; (9) \citet{knispel2015einstein}; (10) \citet{zhu2019mass}; 
(11) \citet{deneva2013goals}; (12) \citet{stovall2019psr}.
\end{itemize}
\end{table}

As discussed by \citet{knispel2015einstein}, the similarity among eMSP$-$HeWDs suggests that they may share a common formation mechanism. Since their discovery, several hypotheses have been proposed, but some of them face limitations. \citet{freire2013direct} proposed that massive WDs may undergo rotation-delayed accretion-induced collapse (AIC) to form NSs. The associated release of gravitational binding energy increases the orbital eccentricity. However, this scenario hardly explains systems in which the inferred NS mass significantly exceeds the Chandrasekhar mass limit for a WD. \citet{wang2023formation} further suggested that direct AIC of super-Chandrasekhar WDs before complete orbital circularization might also leave orbital eccentricity, but this hypothesis has to be tested with simulations. \citet{jiang2015strange} proposed that a phase transition of a massive NS into a strange star could also lead to a sudden mass loss, thereby inducing orbital eccentricity. The model assumes that post-transition MSP masses are distributed between 1.4 and 2.0 ${\rm M}_\odot$, with WD masses of $0.25$–$0.28\,{\rm M}_\odot$. For M10A, the posterior probability of satisfying these constraints is only about $3\%$. On the other hand, \citet{barr2017massive} and \citet{stovall2019psr} argued that assuming a common equation of state, a universal critical density for quark deconfinement, and a constant mass-to-binding-energy conversion fraction, the resulting MSP masses should lie in a relatively narrow range. However, since this conclusion relies on several assumptions, it may not be definitive. \citet{han2021asymmetrical} proposed that asymmetric mass loss driven by thermonuclear shell flashes in proto-white dwarfs can also generate orbital eccentricity. However, as noted by \citet{serylak2022eccentric}, the assumption of instantaneous kicks in this model may be unrealistic. In fact, such flashes typically last for $\sim10^6$–$10^7$ yr (\citealt{sarna2000cooling}). The convective flow resonance model of red giants is another recently proposed formation channel \citep{ginzburg2022eccentric}. In this model, coherent resonance is triggered when the convective eddy turnover time $t_{\rm eddy}$ of the red-giant WD progenitor matches $P_{\rm b}$, eventually leading to the formation of an eccentric system. Moreover, higher (lower) metallicity leads to a longer (shorter) $t_{\rm eddy}$ through its effect on the progenitor's effective temperature. This may explain why some MSPs with $P_{\rm b}$ similar to those of eMSP$-$HeWDs retain nearly circular orbits (e.g., PSR J0732+2314; \citealt{martinez2019discovery}), as their metallicity-dependent $t_{\rm eddy}$ does not match $P_{\rm b}$ and therefore does not trigger resonance. For M10A, however, the low-metallicity environment is expected to favor a relatively short $t_{\rm eddy}$, whereas its $P_{\rm b}$ ($\sim37$ days) is longer than those of known eMSP$–$HeWDs, making this channel less straightforward to apply to M10A.

To date, the model proposed by \citet{antoniadis2014formation} remains the leading explanation for the currently known eMSP$-$HeWDs. In this scenario, the eccentricity arises from interactions with a circumbinary disk formed from material ejected during the hydrogen flashes of the HeWD. The model suggests that HeWD companions should follow the $P_{\rm b}$–$M_{\rm c}$ relation after their evolution as LMXB \citep{tauris1999formation}:
\begin{equation}
\frac{M_{\rm WD}}{{\rm M}_\odot} = \left( \frac{P_{\rm b}}{b} \right)^{1/a} + c,
\label{eq:mwd}
\end{equation}
with $a=5$, $b=10^5$ days, $c=0.110$ for Population II stars. For M10A, this predicts $M_{\rm WD} \sim 0.32\ {\rm M}_\odot$, consistent with the inferred $M_{\rm c}$ at $i \sim 50^\circ$, supporting this scenario as a plausible explanation for the eccentricity of M10A.

In summary, the companion of M10A is most likely a light WD. If the inferred total mass of $2.5(6)\ {\rm M}_\odot$ is not significantly overestimated, the pulsar itself would likely be a massive NS exceeding $2\ {\rm M}_\odot$. Massive NS-light WD systems have already been discovered. One example is PSR~J0348+0432 \citep{antoniadis2013massive, saffer2025lower}, for which $M_{\rm NS}=1.81(4)\, {\rm M}_\odot$ and $M_{\rm WD}=0.154(3)\, {\rm M}_\odot$ \citep{saffer2025lower}, although its orbital properties and evolutionary history differ from those of M10A.

\section{Conclusions}
\label{sec:conclusions}

We have carried out timing observations of the MSP M10A in the GC M10 using FAST. The main results are:

1. M10A is in a wide and eccentric binary system, with orbital eccentricity $e \sim 0.12$, orbital period $P_{\rm b} \sim 37$ days, and projected semi-major axis $x \sim 14$ lt-s. Such a relatively high eccentricity and long orbital period are rare among binary MSPs, suggesting that processes beyond standard binary recycling have shaped its present orbital parameters.

2. We measured a periastron advance rate of $\dot{\omega} = 0.00090(13)$ deg yr$^{-1}$, with no other relativistic effects detected so far. Assuming that the observed periastron advance is entirely relativistic, the total system mass is $M_{\rm tot}=2.5(6)\ {\rm M}_\odot$. From a Bayesian analysis we obtain $M_{{\rm p}} = 2.2^{+0.6}_{-0.5}\ {\rm M}_\odot$ and $M_{{\rm c}} = 0.26^{+0.17}_{-0.03}\ {\rm M}_\odot$ (1$\sigma$ uncertainties), indicating a likely low-mass companion with a massive pulsar. However, this inference remains tentative. First, the uncertainty in $\dot{\omega}$ is still relatively large. Second, the current timing precision is insufficient to precisely measure the proper motion and the parallax, nor to reliably measure the second and higher-order derivatives of the spin frequency, whose contributions may appear as additional $\dot{\omega}$ signals in the ToA residuals, thereby affecting the robustness of the mass inference.

3. The companion of M10A may not have originated from a dynamical exchange, and no optical counterpart was detected at the position of M10A. Therefore, we consider it more likely to be a low-mass WD that formed through standard binary evolution with the MSP and has cooled beyond the detection limits of current observations. The system’s eccentricity may have been induced by long-term gravitational perturbations from nearby bodies, or it may resemble the eMSP$-$HeWD population found outside GCs, suggesting an internally driven eccentricity evolution mechanism independent of GC dynamics.

Further long-term observation campaigns with FAST are already planned to refine the measurement of the proper motion and the parallax, to improve the determination of the periastron advance, and to attempt measurements of more PK parameters, thereby tightening the constraints on the components' masses. Deep optical/UV observations (e.g., with JWST) will be crucial for confirming or excluding a faint WD companion, thereby providing further insight into the evolutionary history of this unusual binary.

\let\internallinenumbers\relax
\nolinenumbers
\begin{acknowledgments}
This work is supported by the National SKA Program of China (No. 2020SKA0120201), the National Key R\&D Program of China (Nos. 2022YFC2205202 and 2020SKA0120100), and the National Natural Science Foundation of China (NSFC; grant Nos. 12373032, 12003047, 11773041, U2031119, 12173052, and 12173053). Both Lei Qian and Zhichen Pan were supported by the Youth Innovation Promotion Association of CAS (id.~2018075, Y2022027 and 2023064), and the CAS "Light of West China" Program. Lijing Shao was supported by the National SKA Program of China (No. 2020SKA0120300), the National Natural Science Foundation of China (No. 12573042), the Beijing Natural Science Foundation (No. 1242018), and the Max Planck Partner Group Program funded by the Max Planck Society. This work made use of the data from FAST (Five-hundred-meter Aperture Spherical radio Telescope) (https://cstr.cn/31116.02.FAST). FAST is a Chinese national mega-science facility, operated by National Astronomical Observatories, Chinese Academy of Sciences. We appreciate all the people from FAST group for their support and assistance during the observations. This work used resources of China SKA Regional Centre prototype (\citealt{an2019ska}, \citealt{an2022status}) funded by Ministry of Science and Technology of the People’s Republic of China and Chinese Academy of Sciences. This work used public data available in the Mikulski Archive for Space Telescopes (MAST) at the Space Telescope Science Institute. The specific observations analyzed can be accessed via \dataset[doi:10.17909/12sm-f327]{https://doi.org/10.17909/12sm-f327}. STScI is operated by the Association of Universities for Research in Astronomy, Inc., under NASA contract NAS5–26555. Support to MAST for these data is provided by the NASA Office of Space Science via grant NAG5–7584 and by other grants and contracts.
\end{acknowledgments}

\vspace{5mm}
\facilities{FAST, HST (WFC3, ACS)}
\software{PRESTO (\citealt{2011ascl.soft07017R}), TEMPO (\citealt{2015ascl.soft09002N}), DRACULA (\citealt{freire2018algorithm}), PySynphot (\citealt{2013ascl.soft03023S}), FEWBODY (\citealt{fregeau2004stellar}, \citealt{2012ascl.soft08011F})}

\bibliography{main}{}
\bibliographystyle{aasjournal}

\end{document}